\documentclass[twocolumn,showpacs,superscriptaddress,amssymb,aps,pra]{revtex4-1}

\usepackage{amsmath,amsfonts,amssymb,amsthm,graphics,graphicx,epsfig,bbm}
\usepackage[colorlinks=true,citecolor=blue,linkcolor=blue,urlcolor=blue]{hyperref}
\usepackage[usenames]{color}
\usepackage{graphicx}
\usepackage{subfigure}
\usepackage{amsmath}
\usepackage{epsfig}
\usepackage{dcolumn}
\usepackage{bm}
\usepackage{color}
\usepackage{epstopdf}
\usepackage{amssymb}
\usepackage{amstext}
\usepackage{latexsym}
\usepackage{hyperref}
\usepackage{psfrag}
\usepackage{xcolor}
\usepackage[normalem]{ulem}
\usepackage{dsfont}
\usepackage{txfonts}
\usepackage{tikz}

\newcommand{\fig}[1]{Fig.~#1}
\newcommand{\eq}[1]{Eq.~(#1)}

\newcommand{\ketc}[1]{\left| #1 \right>} % for Dirac bras

\def\beq{\begin{equation}}
\def\eeq{\end{equation}}
\def\beqa{\begin{eqnarray}}
\def\eeqa{\end{eqnarray}}
\def\bal#1\eal{\begin{align}#1\end{align}}

\begin{document}

\title{Single-photon-driven high-order sideband transitions in an ultrastrongly coupled circuit quantum electrodynamics system}

\author{Zhen Chen}
\thanks{These authors contributed equally to this work.}
\affiliation{Quantum Physics and Quantum Information Division, Beijing Computational Science Research Center, Beijing 100193, China}

\author{Yimin Wang}
\thanks{These authors contributed equally to this work.}
\affiliation{Quantum Physics and Quantum Information Division, Beijing Computational Science Research Center, Beijing 100193, China}

\author{Tiefu Li}
\thanks{litf@tsinghua.edu.cn}
\affiliation{Institute of Microelectronics, Department of Microelectronics and Nanoelectronics and Tsinghua National Laboratory of Information Science and Technology, Tsinghua University, Beijing 100084, China}

\author{Lin Tian}
\thanks{ltian@ucmerced.edu}
\affiliation{School of Natural Sciences, University of California, Merced, California 95343, USA}

\author{Yueyin Qiu}
\affiliation{Quantum Physics and Quantum Information Division, Beijing Computational Science Research Center, Beijing 100193, China}

\author{Kunihiro Inomata}
\affiliation{RIKEN Center for Emergent Matter Science (CEMS), 2-1 Hirosawa, Wako, Saitama 351-0198, Japan}

\author{Fumiki Yoshihara}
\thanks{Current address: National Institute of Information and Communications Technology, 4-2-1,
Nukuikitamachi, Koganei, Tokyo 184-8795, Japan}
\affiliation{RIKEN Center for Emergent Matter Science (CEMS), 2-1 Hirosawa, Wako, Saitama 351-0198, Japan}

\author{Siyuan Han}
\affiliation{Department of Physics and Astronomy, University of Kansas, Lawrence, Kansas 66045, USA}

\author{Franco Nori}
\affiliation{RIKEN Center for Emergent Matter Science (CEMS), 2-1 Hirosawa, Wako, Saitama 351-0198, Japan}
\affiliation{Physics Department, The University of Michigan, Ann Arbor, MI 48109-1040, USA}

\author{J. S. Tsai}
\affiliation{RIKEN Center for Emergent Matter Science (CEMS), 2-1 Hirosawa, Wako, Saitama 351-0198, Japan}
\affiliation{Department of Physics, Tokyo University of Science, Kagurazaka, Shinjuku-ku, Tokyo 162-8601, Japan}

\author{J. Q. You}
\thanks{jqyou@csrc.ac.cn}
\affiliation{Quantum Physics and Quantum Information Division, Beijing Computational Science Research Center, Beijing 100193, China}


\begin{abstract}
We report the experimental observation of high-order sideband transitions at the single-photon level in a quantum circuit system of a flux qubit ultrastrongly coupled to a coplanar waveguide resonator. With the coupling strength reaching 10\% of the resonator's fundamental frequency, we obtain clear signatures of higher-order red- and first-order blue-sideband transitions, which are mainly due to the ultrastrong Rabi coupling.
%in contrast to the previous experiments where high-order sidebands are due to the increase of the pump power.
Our observation advances the understanding of ultrastrongly-coupled systems and paves the way to study high-order processes in the quantum Rabi model at the single-photon level.
\end{abstract}
\date{\today}

\maketitle

\section{Introduction}
\label{sec_intro}
Superconducting quantum circuits exhibit macroscopic quantum coherence (see, e.g., \cite{nakamura_coherent_1999,vion_manipulating_2002,yu_coherent_2002,martinis_rabi_2002,chiorescu_coherent_2004,wallraff_strong_2004,yan_flux_2015}) and can be designed to have exotic properties that cannot be realized or even do not occur in natural atomic systems~\cite{you_atomic_2011}. For instance, the unique geometry of superconducting quantum circuits enables the realization of ultrastrong coupling in qubit-resonator systems with the coupling strength $g$ reaching a considerable fraction of the resonator frequency $\omega_r$: $g/\omega_r \gtrsim 0.1$~\cite{casanova_deep_2010,devoret_circuitqed_2007}.
With current technological advances, ultrastrong coupling has indeed been demonstrated in recent experiments with superconduting flux qubits inductively coupled to superconducting resonators~\cite{niemczyk_circuit_2010, forn-diaz_observation_2010,ultrastrong_2016}. In this ultrastrong-coupling regime, the well-known Jaynes-Cummings model breaks down because the rotating-wave approximation is no longer applicable, and the quantum Rabi model is thus required to describe the energy spectrum and the system dynamics~\cite{rabi_space_1937,braak_integrability_2011}.
Also, the ultrastrong-coupling regime can lead to fast quantum computation schemes~\cite{nataf_protected_2011,romero_ultrafast_2012} and a plethora of interesting quantum optics phenomena~\cite{zueco_qubitoscillator_2009,ashhab_qubitoscillator_2010,nataf_vacuum_2010,ridolfo_photon_2012,
sanchez-burillo_scattering_2014,garziano_multiphoton_2015,Nori-2016,Nori-2017}.

\begin{figure*}[t]
\includegraphics[scale=1]{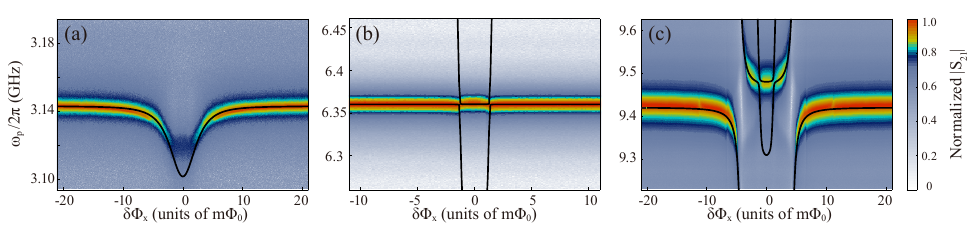}
\caption{
(a) The transmission (i.e. the normalized $|S_{21}|$) spectrum of the $\lambda/2$-mode as a function of the flux bias $\delta \Phi_x$ and probe frequency $\omega_p/2\pi$. The probe power is $P_p\approx-138$~dBm, corresponding to an average photon number of $n_1\approx 1.17$ in the resonator. (b) The spectrum of the $\lambda$-mode with $P_p\approx-135$~dBm and average photon number $n_2\approx 0.24$. (c) The spectrum of the $3\lambda/2$-mode with $P_p\approx-132$~dBm and average photon number $n_3\approx 0.19$. The solid curves are the numerical fit of the spectra with respect to its ground state energy using the full Hamiltonian $H$.}
\label{fig1}
\end{figure*}

Despite its fundamental importance and wide interests, the experimental application of the ultrastrong coupling to quantum information processing remains challenging. It is known that higher-order processes can be achieved with an intense driving field \cite{lev_nat.phy}, but it is difficult to implement these processes with a very weak driving field (i.e, at a few photons level), since it requires large intrinsically-built-in nonlinearity of the system. In contrast to the previous observation of power-enhanced high-order processes by intensifying the driving field, the high-order sideband transitions in our experiment can be realized at the single-photon-driven level and are mainly contributed by the ultrastrong Rabi coupling.
The single-photon-driven first-order sideband transition ($\omega_d = \omega_q \pm \omega_r$) was observed in both strongly and ultrastrongly coupled qubit-resonator systems \cite{wallraff_sideband_2007,forn-diaz_broken_2016}, where $\omega_d$, $\omega_q$ and $\omega_r$ are frequencies of the driving field, qubit and resonator, respectively.
When intensifying the driving field, two-photon-driven first-order sideband transition ($2\omega_d = \omega_q + \omega_r$) was also observed in a strongly coupled qubit-resonator system~\cite{deppe_two-photon_2008,leek_prl(2010)}.
However, in the present experiment, using a suitably-designed ultrastrongly coupled qubit-resonator circuit and a very weak driving field, we are able to resolve up to the third-order sideband transitions $(\omega_d = \omega_q \pm s\omega_r$, $s=0,1,2,3)$ at the single-photon level, where the high-order sideband transitions are mainly due to the ultrastrong Rabi coupling. Also, two-photon-driven second-order sideband transitions $(2\omega_d = \omega_q \pm s\omega_r$, $s=0,1,2)$ can be observed by increasing the power of the driving field. Both the experimental results and the theoretical analyses reveal that the ultrastrong Rabi coupling is the main cause of the high-order sideband transitions presented in our work.

\section{Qubit-resonator circuit}
\label{sec_seup}
Our quantum circuit comprises a superconducting flux qubit inductively coupled to a coplanar waveguide resonator with suitably designed modes (for details, see Appendix A). The superconducting flux qubit consists of four Josephson junctions with three of the junctions designed to be identical and the fourth junction reduced by a factor of $0.6$ in area. The qubit is operated near the optimal flux bias point with an applied external flux $\Phi_x = \delta\Phi_{x}+\Phi_0/2$, where $\Phi_0$ is the magnetic flux quantum and $\delta\Phi_{x}$ is a small offset from the optimal flux bias point $\Phi_0/2$. The qubit Hamiltonian can be written as $H_q = (\epsilon \tau_z + \delta \tau_x)/2 $, where $\delta$ is the quantum tunneling between the local potential wells, $\epsilon = 2 I_p \delta \Phi_x $ is the offset energy induced by the flux bias, with $I_p$ being the maximal persistent current, and $\tau_{z,x}$ are the Pauli operators in the persistent-current basis $\{\ketc{\circlearrowleft}, \ketc{\circlearrowright}\}$~\cite{mooij_josephson_1999}. Below we use the eigenbasis of the qubit $\{\ketc{g}, \ketc{e}\}$ and write the Hamiltonian as $H_{q}=\hbar\omega_{q}\sigma_{z}/2$, with $\omega_{q}=\sqrt{\epsilon^2 +\delta^2}/\hbar$.

To be galvanically connected to the coplanar waveguide resonator~\cite{abdumalikov_vacuum_2008}, the flux qubit shares a common wire (length $34.8$~$\mu$m, width $800$~nm, and thickness $60$~nm) with the resonator's center conductor. The Hamiltonian of the resonator is $H_r = \sum_n \hbar \omega_n (a_n^\dagger a_n + 1/2)$, where $a_n^\dagger $ ($a_n$) is the creation (annihilation) operator of the $n$th resonator modes (i.e., the $n\lambda/2$-mode), and $\omega_n$ is the corresponding resonance frequency. With a transmission measurement, we determine the resonance frequencies of the lowest three modes of the resonator as $\omega_1/2\pi = 3.143 $ GHz,  $\omega_2/2\pi = 6.361 $ GHz, and $\omega_3/2\pi = 9.420 $ GHz. Because of the inhomogeneity of the resonator due to the presence of the qubit, these frequencies are not perfect integer multiples of $\omega_1$~\cite{J. Bourassa_PRA2009}.
Within our parameter range, the $\lambda/2$-mode is dispersively coupled to the flux qubit with a frequency far below the quantum tunneling $\delta$ (i.e., the energy gap at the degeneracy point) of the qubit, and the $\lambda,\, 3\lambda/2$ modes can be tuned to be on resonance with the qubit by adjusting the magnetic flux bias $ \delta \Phi_x $. The dipolar coupling between the qubit and the resonator has the form of
$H_{\textrm{int}} = \sum_n\hbar g_n (a_n^\dagger + a_n )\tau_z$, with ${\hbar}g_n=M I_p I_{r,n}$, where $M$ is the mutual inductance and $I_{r,n}$ is the vacuum center-conductor current of the $n$th resonator mode near the flux qubit.
The qubit is attached to the thin segment of the center conductor in the middle of the resonator, where the current distribution of both the $\lambda/2$- and $3\lambda/2$-mode have antinodes and produces maximum coupling with the qubit. %This design yields ultrastrong coupling between the qubit and the $\lambda/2$-mode. Moreover,
The $\lambda$-mode has a node at this position with nearly negligible coupling to the qubit and will be omitted from our discussion.
The full Hamiltonian of this system is hence $H = H_q + H_r + H_{\textrm{int}}$. The uncoupled states of this system can be expressed as $\ketc{qN_1N_3}$, with $q=\{g,e\}$ representing the qubit eigenstates and $N_n$ being the photon number in the $n$th resonator mode.

\begin{figure*}[!t]
\centering
\includegraphics[scale=1]{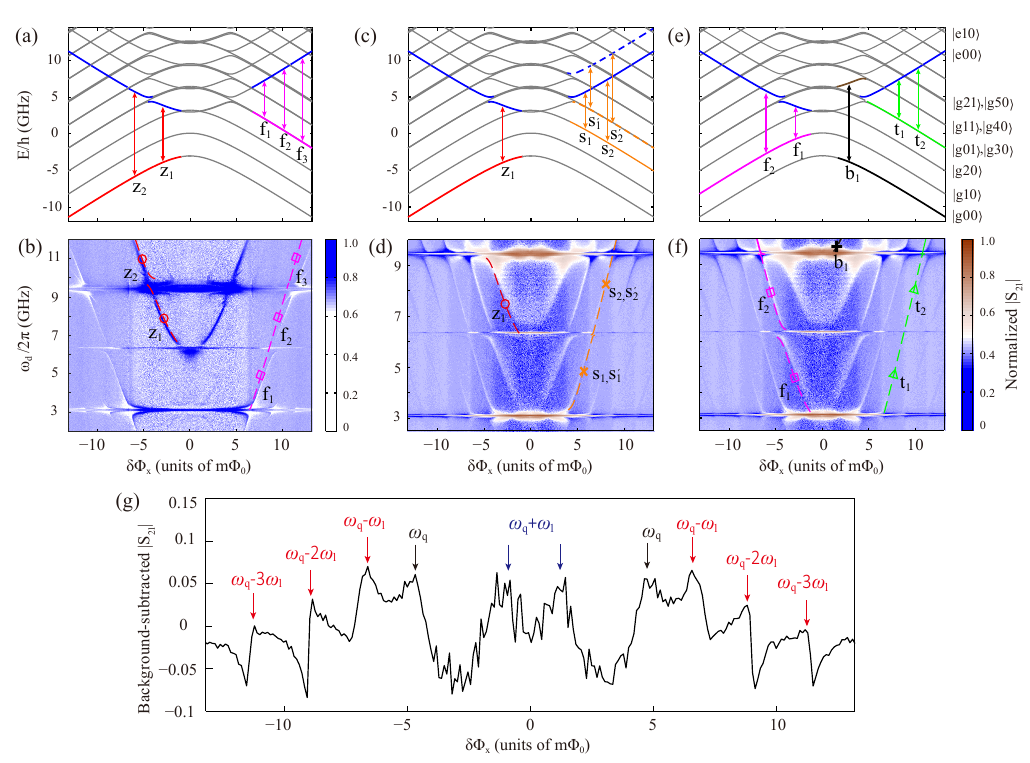}
\caption{
(a), (c), (e) The energy levels of the coupled qubit-resonator system as a function of the flux bias $\delta \Phi_x$ using parameters extracted from \fig{\ref{fig1}}. The energy levels in the dispersive regime are labelled in terms of the uncoupled states $\ketc{qN_1N_3}$.
In (a), the red (magenta) arrows labelled as ${\textrm{z}}_{\textrm{m}}$ (${\textrm{f}}_{\textrm{m}}$) indicate the single-photon-driven zeroth-order qubit (first-order red-sideband) transition.
In (c), the red (orange) arrows labelled as ${\textrm{z}}_{\textrm{m}}$ (${\textrm{s}}_{\textrm{m}}$, ${\textrm{s}}_{\textrm{m}}^{\prime}$) indicate the single-photon-driven zeroth-order qubit (second-order red-sideband and cross-mode sideband) transition.
In (e), the magenta (green) arrows labelled as ${\textrm{f}}_{\textrm{m}}$ (${\textrm{t}}_{\textrm{m}}$) indicate the single-photon-driven first-order (third-order) red-sideband transition and the black arrow labelled as ${\textrm{b}}_{1}$ indicates the single-photon-driven first-order blue-sideband transition.
(b), (d), (f) The transmission (normalized $|S_{21}|$) spectrum as a function of the flux bias $\delta \Phi_x$ and driving frequency $\omega_d/2\pi$.
In (b), for probe frequency of the $3\lambda/2$-mode with probe power $P_p\approx-132$ dBm (average photon number $n_3\approx 0.19$). The red-circle lines are due to the single-photon-driven zeroth-order qubit transition $\ketc{g00} \leftrightarrow \ketc{e00}$, corresponding to the red arrows in (a). The magenta-square lines are due to the single-photon-driven first-order red-sideband transition $\ketc{g01} \leftrightarrow \ketc{e00}$, corresponding to the magenta arrows in (a).
(d) and (f), for probe frequency of the $\lambda/2$-mode with probe power $P_p\approx-128$ dBm (average photon number $n_1\approx 11.7$).
In (d), the red-circle (orange-cross) lines are due to the single-photon-driven zeroth-order qubit (second-order red-sideband and cross-mode sideband) transition.
In (f), the magenta-square (green-triangle) lines are the single-photon-driven first-order (third-order) red-sideband transition and the black-cross denoted short line indicates the single-photon-driven first-order blue-sideband transition. The power of the driving field on the resonator is $P_d\approx-92$ dBm for (b) and $P_d\approx-97$ dBm for (d) and (f). The effective field to drive the flux qubit is much reduced in the dispersive regime of the qubit-resonator system and the resulting average photon number in the resonator is far less than one (see Fig.~\ref{figs3} in Appendix B). (g) A linecut of the background-subtracted transmission spectrum in (d) at $\omega_d/2\pi$ = 9.78 GHz, where the first-order blue sideband transitions purely due to the counter-rotating terms are marked with the blue arrows.}
\label{fig2}
\end{figure*}

\section{Transmission spectra}
\label{sec_transmission}
In measuring the transmission spectrum of the qubit-resonator system, we apply a single probe source of frequency $\omega_p$ to the resonator via a vector network analyzer and measure the resonator output at the probe frequency. In the experiment, a low-power probe source is used to avoid producing any appreciable effects on the qubit-resonator system.
Figure~\ref{fig1} shows the (color-coded) transmission spectra in the neighborhood of the resonance frequencies of the $\lambda/2$-, $\lambda$- and $3\lambda/2$-mode, respectively. The measured spectral structures in these plots correspond to the transition frequencies between the ground and excited states. To find the magnitudes of the coupling strength $g_{n}$, we calculate the eigenstates of the full Hamiltonian $H$ numerically and fit the measured data to the calculated energy splittings. The calculated transition frequencies are plotted as black curves in Fig.~\ref{fig1} with $g_1/2\pi = 306$ MHz, $g_2/2\pi = 5$ MHz, and $g_3/2\pi = 521$ MHz. These coupling strengths give the coupling ratios $g_1/\omega_1 = 9.74\%$,  $g_2/\omega_2 = 0.08\%$, and $g_3/\omega_3 = 5.53\%$.

\section{Single-photon-driven high-order sideband transitions}
\label{sec_1psideband}
With ultrastrong Rabi coupling, single-photon-driven high-order sideband transitions can be observed in transmission spectroscopic measurements by using a weak pump field at frequency $\omega_d$ to drive the qubit through the resonator. The pump Hamiltonian has the form $H_d=\Omega_{d,q}\cos{(\omega_d t)}\tau_z$ in the persistent-current basis, with $\Omega_{d,q}$ being the driving strength. A separate probe field with its frequency fixed at the resonance frequency of one of the resonator modes is applied to demonstrate the spectroscopic response of the coupled qubit-resonator system in the presence of the pump field. The transmission spectra are measured by monitoring the amplitude and the phase of the transmitted probe tone~\cite{schuster_ac_2005,niemczyk_circuit_2010}. In Fig.~\ref{fig2}, we show the measured transmission spectra of the probe field at a probe frequency $\omega_{3}$ of the $3\lambda/2$-mode [Fig.~\ref{fig2}(b)], and at a probe frequency $\omega_{1}$ of the $\lambda/2$-mode [Fig.~\ref{fig2}(d, f)], respectively.
Note that Figs.~\ref{fig2}(b), \ref{fig2}(d) and \ref{fig2}(f) present some vertical banding as a function of the flux bias, and the data in the measured transmission spectra become noisier at low flux bias. In fact, when the flux bias $\delta \Phi_x$ approaches the degenerate point of the qubit, the resonance frequencies of the resonator shift [see Figs.~\ref{fig1}(a) and \ref{fig1}(c)], due to the ultrastrong coupling between the qubit and the resonator. This makes the measured signals weaker at low flux bias than at high flux bias, thus yielding noisier data at low flux bias and also the appearance of the vertical banding versus the flux bias.

To identify each transition, in Figs.~\ref{fig2}(a), ~\ref{fig2}(c) and ~\ref{fig2}(e), we show the energy levels of the total qubit-resonator system as a function of the flux bias $\delta \Phi_x$ with the colored arrows labeling the corresponding sideband transitions in the presence of the pump field. In these plots, besides the main peaks at the pump frequency $\omega_d = \omega_q$ (see the red-circle denoted lines) that correspond to the direct single-photon transition $\ketc{g00}\leftrightarrow\ketc{e00}$ (denoted by ${\textrm{z}}_1$ and ${\textrm{z}}_2$), we also observe single-photon-driven high-order sideband transitions due to the ultrastrong Rabi coupling.
With a Schrieffer-Wolff transformation~\cite{zueco_qubitoscillator_2009,bravyi_schriefferwolff_2011}, we can identify the single-photon-driven higher-order sideband transitions in the dispersive regime when the qubit frequency is far off resonance from the resonator frequencies, and compare the transitions with the measured spectra (see detailed discussions in Appendix B).

In Fig.~\ref{fig2}(b), spectral features are observed at the pump frequency $\omega_d = \omega_q - \omega_3$, as indicated by the magenta-square denoted lines
 ${\textrm{f}}_{1}$, ${\textrm{f}}_{2}$ and ${\textrm{f}}_{3}$. With a pump field on the $\sigma_{z}$-component of the qubit, one can have effective qubit resonances at $\omega_{q} \mp s\omega_{r}$, with $s$ being an integer~\cite{deng_observation_2015}. In the dispersive regime, due to the combination of single-photon pumping and the qubit-resonator interaction, the observed spectral lines corresponding to the single-photon-driven first-order red-sideband transition $\ketc{g01}\leftrightarrow\ketc{e00}$, are enabled by shifting the driving frequency from $\omega_{q}$ to $\omega_{q}-\omega_{r}$ .

The effective Hamiltonian derived with the Schrieffer-Wolff transformation is $H_{\textrm{eff}} = R_{3}^{(1)} (\sigma_{+} a_{3}+a_{3}^{\dag} \sigma_{-})$, with the coupling coefficient $R_{3}^{(1)} = -\Omega_{d,q} \cos \theta g_{3}/\Delta_{3}^{-} $, where $\Delta_{3}^{\pm}=\omega_{q}\pm\omega_{3}$ and $|\Delta_{3}^{\pm}|\gg g_{3}$. The transitions induced by these effective couplings are also labelled in \fig{\ref{fig2}}(a).

Single-photon-driven higher-order sideband transitions are shown in Figs.~\ref{fig2}(c)-\ref{fig2}(f), where the transmission spectrum is measured at the frequency of the $\lambda/2$-mode with the probe tone being 4 dB higher and the driving tone being 5 dB lower than the signals used in \fig{\ref{fig2}}(b). Here single-photon-driven red-sideband transitions [see the arrows in Figs.~\ref{fig2}(c) and \ref{fig2}(e)] up to the third-order ($\omega_{d}=\omega_q - s \omega_1$, with $s=1,2,3$) are observed, as indicated by the corresponding colored lines in Figs.~\ref{fig2}(d) and \ref{fig2}(f). The peaks at $\omega_{d}=\omega_q - \omega_1$ (labelled by ${\textrm{f}}_1$ and ${\textrm{f}}_2$) are dominated by the single-photon-driven first-order red-sideband transition, as analyzed above. The peaks at $\omega_{d}=\omega_q - 2\omega_1$ are a mixture of the single-photon-driven second-order red-sideband transition $\ketc{g20}\leftrightarrow\ketc{e00}$  (labelled by ${\textrm{s}}_1$ and ${\textrm{s}}_2$) induced by the effective coupling $H_{\textrm{eff}} = R_{1}^{(2)}  (\sigma_{+} a_{1}^{2}+a_{1}^{\dag 2} \sigma_{-})$ and a cross-mode sideband transition $\ketc{g01}\leftrightarrow\ketc{e10}$  (labelled by ${\textrm{s}}'_{1}$ and ${\textrm{s}}'_{2}$) by the effective coupling $H_{\textrm{eff}} =  R^{(2)}_{\bar{1}3}(\sigma_{+} a_{1}^{\dag} a_{3}+a_{3}^{\dag}a_{1} \sigma_{-})$. These two transitions have comparable frequencies because the cross sideband frequency $\omega_{3}-\omega_{1}\approx 2\omega_{1}$ in our device. Expressions for the coupling constants $R_{1}^{(2)}$ and $R^{(2)}_{\bar{1}3}$ can be found in Appendix B. The single-photon-driven third-order red-sideband transition $\ketc{g30}\leftrightarrow\ketc{e00}$  (labelled by ${\textrm{t}}_{1}$ and ${\textrm{t}}_{2}$) is observed at a pump frequency $\omega_{d}=\omega_q - 3\omega_1$. It originates from the effective coupling $H_{\textrm{eff}} = R_{1}^{(3)}  (\sigma_{+} a_{1}^{3}+a_{1}^{\dag 3} \sigma_{-})$ with $R_{1}^{(3)} =2 \Omega_{d,q}\cos\theta {(g_1/\Delta_1^-)}^2(g_1/\Delta_1^+) /3$, which reveals that this transition depends on the counter-rotating terms. In addition to the red-sidebands, the single-photon-driven first-order blue-sideband is also observed, which represents the transition $\ketc{g00}\leftrightarrow\ketc{e10}$ as indicated by the black arrow and the black-cross line  (labelled by ${\textrm{b}}$) in Figs.~\ref{fig2}(e) and \ref{fig2}(f).
This transition is induced by the effective coupling $H_{\textrm{eff}} =  B_{1}^{(1)} (\sigma_{+} a_{1}^{\dag}+a_{1} \sigma_{-})$ with $B_{1}^{(1)} = -\Omega_{d,q} \cos \theta g_1/\Delta_1^+ $, which is purely due to the counter-rotating terms. In Fig.~\ref{fig2}(g), we also show a linecut of the transmission spectrum in Fig.~\ref{fig2}(d) extracted at $\omega_d/2\pi$ = 9.78 GHz, so as to clearly exhibit this transition.

As indicated in Figs.~\ref{fig2}(a), \ref{fig2}(c) and \ref{fig2}(e), all sideband transitions observed in Figs.~\ref{fig2}(b), \ref{fig2}(d) and \ref{fig2}(f) involve the qubit states from $\ketc{g}$ to $\ketc{e}$, as well as the Fock states of the resonator from $\ketc{0}$ to $\ketc{n}$, $\ketc{n}$ to $\ketc{0}$, or $\ketc{n}$ to $\ketc{n'}$, where $n$, $n'$ = 1, 2, $\cdots$. This implies that excited-state populations of the resonator are needed for these sideband transitions. In Fig.~\ref{fig2}(b), where the first-order sideband transitions are observed, the average photon number is estimated to be $n_3\approx 0.19$ for the probe field in resonance with the $3\lambda/2$-mode of the resonator. In Figs.~\ref{fig2}(d) and \ref{fig2}(f), where up to the third-order sideband transitions are observed, the average photon number is then estimated to be $n_1\approx 11.7$ for the probe field in resonance with $\lambda/2$-mode of the resonator. These given values of the average photon number of the probe field indeed reveal the possible population of the excited states in the resonator. Also, thermal fields may contribute to the excited-state population of the resonator, but they are not as important as the probe field in our experiment (see the discussions in Sec.~VI as well).

Note that the magnitude of the $s$th-order single-photon-driven sideband transition has linear dependence on the qubit-driving strength $\Omega_{d,q}$ as that of the zeroth-order process (i.e., the direct transition of the qubit states by the pumping field). Also, it has power-law dependence on the coupling ratio as $(g_{n}/\Delta_{n}^{\pm})^{s}$, as explained in detail in Appendix B. Thus, at the single-photon level to observe the zeroth-order process by using a very weak pumping field, the high-order processes can also be demonstrated with a strong enough qubit-resonator coupling.  This is the case in our experiment, where the qubit-resonator coupling is ultrastrong and the high-order processes are observed at the quantum limit. As shown in Appendix B, in the dispersive regime, the average photon number of the driving field is even much less than one in the resonator. Therefore, the transmission spectra in Fig.~\ref{fig2} reveal that the effects are mostly connected to the ultrastrong Rabi coupling of the qubit-resonator system.

It is worthwhile to mention that the numerical fittings do not overlap perfectly with the slope of the experimental spectra for all regions of the flux biases, as shown in Figs.~\ref{fig2} and \ref{fig3}. For large flux biases, i.e. $|\delta \Phi_x| \gtrsim$ 8 m$\Phi_0 $, small deviations appear between the fitting lines and the measured spectra. For these, there are two reasons. One is the presence of higher resonator modes in the real system, which cannot be captured numerically due to computational limitations. The other reason is that the two-level approximation for the flux qubit is not good enough at larger flux biases. Close to the degenerate point, the two-level approximation works well and we have a nearly perfect two-level system that can be well described by Pauli operators. However, away from the degenerate point, higher levels start to affect the system's dynamics.

\section{Two-photon-driven high-order sideband transitions}
\label{sec_2psideband}
\begin{figure}[!b]
\includegraphics[scale=1.1]{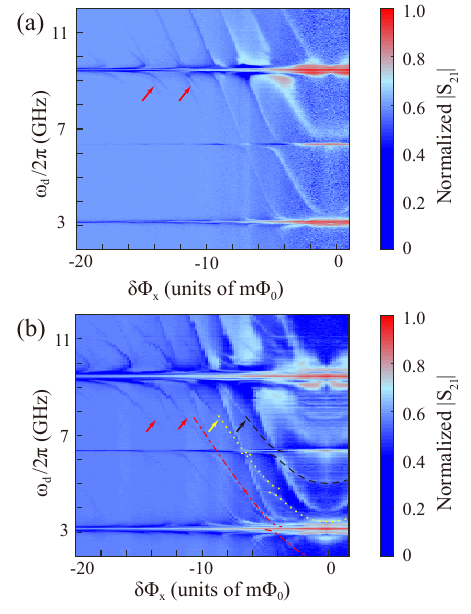}
\caption{
The transmission spectroscopy (normalized $|S_{21}|$) as a function of the flux bias $\delta \Phi_x$ and driving frequency $\omega_d/2\pi$ probed at the frequency of $\lambda/2$ mode with power of $P_p\approx-133$ dBm, which contributes an average photon number $n_1\approx 3.34$ into the resonator. With the driving power increased to (a) $P_d\approx-92$dBm and (b) $P_d\approx-82$dBm, the two-photon-driven sideband transitions $(2\omega_d = \omega_q \pm s \omega_1)$ become visible, as indicated by the yellow arrows for the two-photon qubit transitions with $s=0$, the red arrows for the two-photon first- and second-order red-sideband transitions, and the black arrows for the two-photon first-order blue-sideband transitions. For clarity, here we only show in \fig{\ref{fig3}}(b) the numerical fitting of the two-photon-driven first-order sideband transitions as labelled by the yellow, red, and black curves for $2\omega_d = \omega_q$, $2\omega_d = \omega_q - \omega_1$, and $2\omega_d = \omega_q + \omega_1$, respectively. }
\label{fig3}
\end{figure}
To further illustrate that our high-order sideband transitions are induced by the ultrastrong coupling rather than the driving power, here we show the spectroscopy measurement with the driving power increased to $P_d\approx-92$ dBm in Fig.~\ref{fig3}(a) and $P_d\approx-82$ dBm in Fig.~\ref{fig3}(b), respectively.
In addition to the single-photon-driven first-, second- and third-order sideband transitions, we can clearly resolve two-photon-driven high-order red-sideband transitions at $2\omega_d = \omega_q - s \omega_1$, with $s=0,1$ in Fig.~\ref{fig3}(a) and $s=0,1,2$ in Fig.~\ref{fig3}(b). Also, two-photon-driven first-order blue-sideband transition at $2\omega_d = \omega_q + \omega_1$ is visible.
The observation of two-photon-driven sideband processes with $s=0,1,2$ rather than single-photon-driven higher-order processes with $s=4,5,6, \dots$ by intensifying the driving field further reveal that the single-photon-driven high-order sideband transitions in Fig.~\ref{fig2} depend more significantly on the qubit-resonator coupling strength. This indicates that the larger the coupling strength is, the easier to observe the single-photon-driven higher-order sideband transitions with a relatively weak driving field.

As shown in Fig.~\ref{fig3}, the single-photon-driven high-order sideband transitions can also be induced by intensifying the driving field, but when increasing the driving power, multi-photon-driven high-order sideband transitions (i.e., $m\omega_d = \omega_q \pm s\omega_n$, with $m=1,2,3,\dots$) will also occur (in Fig.~\ref{fig3}, two-photon-driven high-order sideband transitions $2\omega_d = \omega_q \pm s\omega_1$ indeed occurred when intensifying the driving field). Moreover, in addition to the complicated spectral features, the peaks related to the single-photon-driven high-order sideband transitions become blurred by increasing the driving power [see Fig.~\ref{fig3}(b)]. Therefore, the existence of ultrastrong Rabi coupling is a necessary condition to observe the high-order sideband transitions in Fig.~\ref{fig2} at the quantum limit of single-photon level.

\section{Discussions and conclusions}
\label{sec_con}
Higher-order processes at the few-photon level can play an important role in quantum information processing. In principle, the generation of higher-order processes requires large nonlinearity in the system, which can be either intrinsically-built-in or externally-induced. When driven strongly, the system generates nonlinearity externally as in classical nonlinear optics. Therefore, high-order sideband transitions may be demonstrated using strong driving on a system with small coupling strength.  However, in the present work, we demonstrate that the higher-order processes can also be produced in the weak limit of driving field (i.e., at the few-photon level) on the qubit, where the nonlinearity mainly comes from the intrinsic properties of the ultrastrongly coupled system. In fact,
it is clearly analyzed in Appendix B that in the dispersive limit, the magnitudes of the single-photon-driven $s$th-order sideband processes have both a linear dependence on the amplitude of the driving field on the qubit and a power-law dependence on the coupling ratio as $(g_{n}/\Delta_{n}^{\pm})^{s}$. When the coupling strength is not strong enough, the higher-order sideband transitions can be very weak and become hard to be implemented with a weak driving field on the qubit. Therefore, our experimental observation of the high-order sideband transitions at the single-photon-driven level reveals the importance of the ultrastrong coupling.

When applying driving microwave powers to the resonator at the level of -90dBm or higher, though the effective driving field on the qubit is weak (see Appendix B), there can be a considerable amount of heat generated in the attenuator anchored to the mixing chamber stage of the dilution refrigerator. Even when the temperature of the plate itself does not change, hot electrons may radiate thermal fields which could drive the resonator and create excitations that would explain the appearance of higher-order sidebands at higher powers. In fact, similar techniques of increasing the effective resonator temperature were used in other experiments (see, e.g., \cite{thermal Fink2009}) as a way to study higher-level transitions in qubit-resonator systems. However, in our experiment, only single-photon-driven high-order sideband transitions are observed in the experimental results shown in Fig.~\ref{fig2}. If the thermal fields were important, more sideband transitions driven by multiple photons, as seen in Fig.~\ref{fig3}, would appear in the case of Fig.~\ref{fig2}. Actually, such features do not occur in Fig.~\ref{fig2}. Moreover, our spectra in Fig.~~\ref{fig2} do not show the many transitions between dressed states, as observed in \cite{thermal Fink2009} owing to the thermal excitations. These indicate that the thermal fields do not play an appreciable role in our experiment.

In Fig.~\ref{fig2}(b), there are tiny features for some other transitions, which appear in the near-resonance regions of the resonator modes. Moreover, the spectra in the on-resonance regions become more complicated owing to the very large driving power. This is beyond the scope of studies in the present work, because we only focus on the dispersive regimes. It will be extensively studied in our future work.

In conclusion, we have observed high-order sideband transitions in an ultrastrongly coupled qubit-resonator system at the single-photon level. These transitions, including red-sideband transitions up to the third order and first-order blue-sideband transition, are mainly induced by the ultrastrong Rabi coupling rather than the strong pump power. Also, we demonstrated the two-photon-driven high-order sideband transitions in this ultrastrongly coupled system by intensifying the driving field.
Our results provide better understanding of high-order processes in the ultrastrong Rabi model at the single-photon-driven level.

\section*{Acknowledgements}

We thank P.-M. Billangeon and D.-K. Zhang for valuable discussions and K. Kusuyama for technical assistance. This work is supported by the National Key Research and Development Program of China Grant No.~2016YFA0301200, the NSAF Grant Nos.~U1330201 and U1530401, the National Basic Research Program of China Grant Nos. 2014CB848700 and 2014CB921401, and the NSFC Grant No.~91421102. Y.M.W. is partly supported by the NSFC Grant No.~11404407, the Jiangsu NSF Grant No.~BK20140072 and China Postdoctoral Science Foundation Grant Nos. 2015M580965 and 2016T90028. L.T. is partially supported by the National Science Foundation under Award No. NSF-DMR-0956064 and thanks CSRC for hospitality during her visit. S.H. acknowledges support from NSF (Grant No. DMR-1314861) and thanks CSRC for hospitality. F.N. was partially supported by the RIKEN iTHES Project, MURI Center for Dynamic Magneto-Optics via the AFOSR Award No.~FA9550-14-1-0040, the Japan Society for the Promotion of Science (KAKENHI), the IMPACT program of JST,
JSPS-RFBR grant No.~17-52-50023, CREST grant No.~JPMJCR1676, and the Sir John Templeton Foundation. J.S.T. is partially supported by the Japanese Cabinet Office's ImPACT project.

%%%%%%%%%%%%%%%%%%%%%%%%%%%%%%%%%%%%%%%%%%%%%%%%%%%%%%%%%%%%%%%%%%%%%%%%%%%%%%%%%
%%%%%%%%%%%%%%%%%%%%%%%%%%%%%%%%%%%%%%%%%%%%%%%%%%%%%%%%%%%%%%%%%%%%%%%%%%%%%%%%%

\renewcommand{\theequation}{A-\arabic{equation}}
% redefine the command that creates the equation no.
\setcounter{equation}{0}  % reset counter

\renewcommand{\thefigure}{A\arabic{figure}}
% redefine the command that creates the figure no.
\setcounter{figure}{0}  % reset counter
%\begin{widetext}
\section*{Appendix A: Fabrication and device characterization}
\label{sec Appendix1}
\begin{figure*}[t]
\includegraphics[scale=1.2]{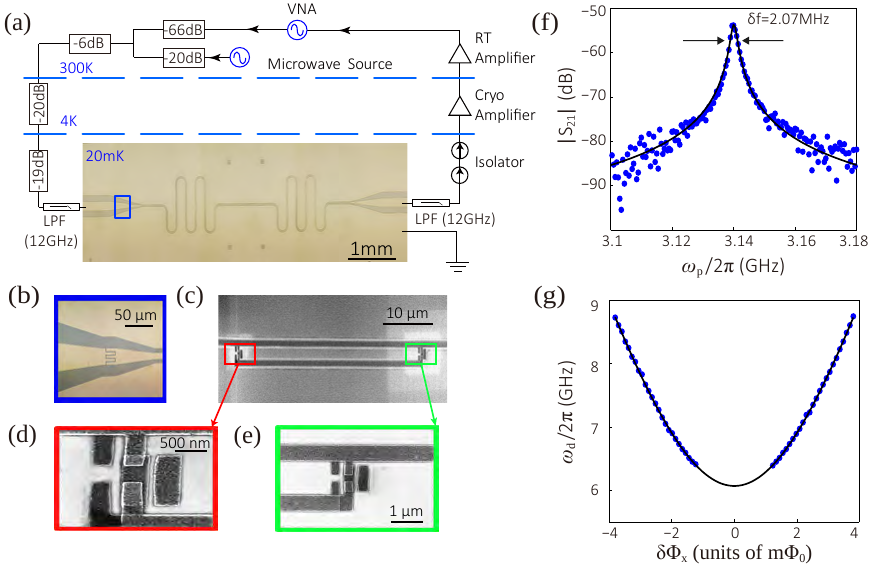}
\caption{
(a)~
Optical image of the superconducting $\lambda/2$ coplanar waveguide resonator and schematic representation of the experimental setup. The transmission through the cavity at frequency $\omega_p$ is measured using a vector network analyser (VNA). A second microwave signal at frequency $\omega_d$ is used for qubit spectroscopy measurement. The input signal is attenuated and filtered at various temperature stages and coupled into the resonator through the capacitor.  Two isolators and low-pass filters (LPF) are used to protect the sample from the cryo-amplifier's noise.
(b)~
Optical image of one of the two identical coupling capacitors of the resonator, as indicated by the blue rectangle area in (a).
(c)~
Scanning electron microscope (SEM) image of the galvanically-connected flux loop (the red and green rectangle areas are shown in (d) and (e), respectively). The shared arm between the flux qubit and the resonator's center line is $34.8$~$\mu$m long and $800$~nm wide.
(d)~
SEM image of the left two Josephson junctions in the flux qubit loop.
(e)~
SEM image of the right two Josephson junctions in the flux qubit loop.
(f)~
Transmission spectra of the $\lambda/2$ resonator mode, which is measured at $20$~mK. The black-continuous line shows the Lorentzian fit to the transmission power spectrum. The resonance frequency $\omega_1/2\pi=3.143$~GHz and FWHM of $2.07$~MHz are obtained by the fitting.
(g), Qubit transition frequency $\omega_q$ from spectroscopy measurement versus relative magnetic flux bias $\delta \Phi_x$. The data is recorded at a probe power $P_p\approx -143$ dBm (corresponding to an average photon number $n_1\approx 0.19$ in the resonator). The probe frequency is equal to the frequency of the $\lambda/2$-mode. The black-continuous line represents a numerical fit to the qubit Hamiltonian $H_q$ yielding the parameters $\delta/h = 6$~GHz and $I_p =265$~nA.}
\label{fig_sfig1}
\end{figure*}

The coplanar waveguide resonator is fabricated by electron-beam lithography and reactive ion etching on a 3-inch thermally oxidized silicon wafer covered with a $50$~nm thick d.c.-magnetron sputtered niobium film [Fig.~\ref{fig_sfig1}(a)]. The center conductor of the resonator is $20$~$\mu$m wide and its gap to the ground plane is $11.6$~$\mu$m, so that a $50$~$\Omega$ characteristic impedance is obtained. The resonator with a length of $16$~mm is defined by two identical interdigital coupling capacitors [Fig.~\ref{fig_sfig1}(b)] with a numerically simulated capacitance of about $7$~fF. In the middle of the center conductor, a $100$~$\mu$m long niobium film is replaced by an aluminium strip which connects to the flux qubit [Fig.~\ref{fig_sfig1}(c)]. The aluminum part is fabricated using electron-beam lithography and Al/AlOx/Al shadow evaporation techniques. The thickness of the bottom and top layer is $25$~nm and $35$~nm. For the flux qubit, three of the Josephson junctions have an area $500$~nm$\times 400$~nm and the other junction is $205$~nm $\times400$~nm, reduced by a factor of $0.6$ [Fig.~\ref{fig_sfig1}(d) and \ref{fig_sfig1}(e)]. The area of the qubit loop is $34.8$~$\mu$m $\times 3.3$~$\mu$m. The $34.8$~$\mu$m-long shared arm generates ultrastrong qubit-resonator coupling.

The quantum circuit is characterized at a temperature of $20$~mK in a BlueFors LD-400 dilution refrigerator. A transmission measurement is conducted to measure the resonator properties in the low-power limit of the probe field. As shown in Fig.~\ref{fig_sfig1}(f), the $\lambda/2$-mode has a resonant frequency of $3.143$~GHz and full width at half maximum (FWHM) of $2.07$~MHz. By a transmission spectroscopy measurement [Fig.~\ref{fig_sfig1}(g)], the dependence of the qubit frequency on the applied external flux is obtained.

\section*{Appendix B: High-order sideband transitions in the dispersive limit}
\subsection{The ultrastrongly coupled system under a resonator driving}
In the persistent-current basis $\{\ketc{\circlearrowleft}, \ketc{\circlearrowright}\}$ of the superconducting flux qubit, the dynamics of the coupled qubit-resonator system is governed by the Hamiltonian
\bal
\label{eq_hfull1}
H=&\frac{1}{2} (\epsilon \tau_z + \delta \tau_x) + \sum_{n=1,3}\left[ \hbar \omega_{n} a_{n}^{\dagger }a_{n} + \hbar g_n (a_n^\dagger + a_n ) \tau_z \right].
\eal
When applying a pump field of frequency $\omega_d $ to the coplanar waveguide resonator, the total Hamiltonian becomes
\bal
H_t &= H + H_d, \nonumber \\
%\label{eq_hfull2}
H_d &= \sum_{n=1,3} \Omega_{r,n}\cos{\omega_d t}\, (a_{n}^{\dagger } + a_{n}),
\label{eq_hfull3}
\eal
with $\Omega_{r,n}$  being the Rabi frequency of the driving field on the $n$th mode of the resonator.

To analyze the effect of the driving field on the flux qubit, we displace the field operator using a time-dependent displacement operator
\bal
D(t) = \exp{X(t)},~~ X(t) = \sum_{n=1,3} \left[\alpha_n(t) a_{n}^{\dagger} - \alpha_n^*(t) a_{n}\right].
\eal
The displaced Hamiltonian now reads
\bal
H'=& D^\dagger (t) H_t D(t)  - i D^\dagger (t) \partial_t D(t) \nonumber \\
=& H_t + [H_t,X(t)] + \frac{1}{2}[[H_t,X(t)],X(t)] + ... \nonumber \\
&- i D^\dagger (t) D(t) \partial_t X(t) \nonumber \\
= & \frac{1}{2} (\epsilon \tau_z + \delta \tau_x) + \sum_{n=1,3} \left[ \hbar \omega_{n} a_{n}^{\dagger }a_{n} + \hbar g_n (a_n^\dagger + a_n ) \tau_z \right] \nonumber \\
&+ \sum_{n=1,3} g_n \left[\alpha_n(t) +  \alpha^*_n(t)\right]\, \tau_z,
\label{eq_hdfull}
\eal
where $\alpha_n(t)$ is chosen to be
\bal
\alpha_n(t) = -\frac{\Omega_{r,n}}{2} \left( \frac{1}{\omega_n-\omega_d} e^{-i \omega_d t} +  \frac{1} {\omega_n+\omega_d} e^{i \omega_d t} \right),
\label{eq_alphat1}
\eal
which satisfies the following equation
\bal
\partial_t \alpha_n(t) = -i \omega_n \alpha_n(t) - i \Omega_{r,n}\cos{\omega_d t}.
\eal
Then the last term $g_n \left[\alpha_n(t) +  \alpha^*_n(t)\right]\, \tau_z$ in \eq{\ref{eq_hdfull}}, which represents the effective driving filed on the qubit, can now be written as $\Omega_{d,q} \cos{\omega_d t }\,\tau_z$, with
\bal
\Omega_{d,q} & = \Omega_{q,1} + \Omega_{q,3}, \nonumber \\
\Omega_{q,n} &= g_n \,\Omega_{r,n} \left( \frac{1}{\omega_d-\omega_n } -  \frac{1} {\omega_n+\omega_d}\right),
\label{eq_rabiatom}
\eal
where $|\Omega_{q,n}|$ being the effective Rabi frequency of the driving field on the flux qubit via the $n$th-mode of the resonator.
When the linewidth of the cavity mode is considered, $\omega_n$ in \eq{\ref{eq_alphat1}} is replaced by $\omega_n+i\kappa_n/2$, where $\kappa_n$ is the total photon damping rate of the $n$th cavity mode, which can be written as the sum of the individual contributions from the external and internal channels, i.e. $\kappa_{n} =\kappa_{n,{\rm in}} + \kappa_{n,{\rm ex}}$ \cite{aspelmeyer_cavity_2014}. Here in our system, the loss rate associated with
the waveguide-resonator interface $\kappa_{n,{\rm ex}}$ is much larger than the loss rate inside the resonator  $\kappa_{n,{\rm in}}$, i.e. $\kappa_{n,{\rm ex}} \approx\kappa_n \gg\kappa_{n,{\rm in}} $, and thus we ignore $\kappa_{n,{\rm in}}$ for the following numerical estimation. We measured the total cavity photon loss rates of $\kappa_1/2\pi \approx 2.07$ MHz, $\kappa_2/2\pi \approx 9.90$ MHz and $\kappa_3/2\pi \approx 18.01$ MHz for the $\lambda/2-$, $\lambda-$ and $3\lambda/2-$mode, respectively.

In this case, the qubit driving term becomes
\begin{widetext}
\bal
 2\,g_n \,\Omega_{r,n}\,\omega_n \left[ \frac{\omega_d^2-\omega_n^2-\kappa_n^2/4}{{(\omega_d^2-\omega_n^2-\kappa_n^2/4)}^2+\omega_d^2\kappa_n^2} \cos \omega_d t +  \frac{\omega_d\kappa_n}{{(\omega_d^2-\omega_n^2-\kappa_n^2/4)}^2+\omega_d^2\kappa_n^2} \sin\omega_d t  \right],
\label{eq_hqkappa}
\eal
\end{widetext}
where the second term in \eq{\ref{eq_hqkappa}} can be safely ignored since $\kappa_n \ll \{\omega_n, \omega_d\}$, and $\Omega_{q,n}$ then reads
\bal
\Omega_{q,n}= 2\,g_n \,\Omega_{r,n}\,\omega_n \frac{\omega_d^2-\omega_n^2-\kappa_n^2/4}{{(\omega_d^2-\omega_n^2-\kappa_n^2/4)}^2+\omega_d^2\kappa_n^2}.
\eal
\begin{figure}[!b]
\centering
\includegraphics[scale=1.1]{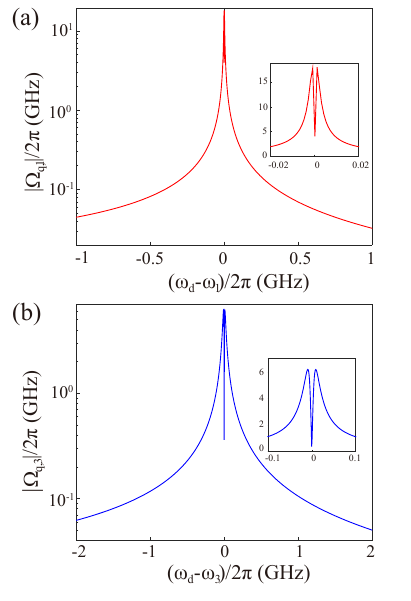}
\caption{
Rabi frequency of the driving field on the flux qubit for (a) the $\lambda/2$-mode and (b) $3\lambda/2$-mode. Each inset is a zoom of the near-resonance region between the driving field and the corresponding mode. }
\label{figs2}
\end{figure}
Since $\Omega_{r,n}$ can be directly expressed in terms of the driving power as \cite{steck_quantum_2015}
\bal
\Omega_{r,n} = \sqrt{\frac{ P_d \kappa_{n,{\rm ex}}}{2\, \hbar \omega_n}},
\label{eq_omgr}
\eal
$\Omega_{q,n}$ can then be written as
\bal
\Omega_{q,n} = g_n \,\sqrt{\frac{2\, P_d \,\omega_n \kappa_{n,{\rm ex}}}{\hbar}} \frac{\omega_d^2-\omega_n^2-\kappa_n^2/4}{{(\omega_d^2-\omega_n^2-\kappa_n^2/4)}^2+\omega_d^2\kappa_n^2}.
\label{eq_omgq}
\eal

Note that the effective Rabi frequency $|\Omega_{q,n}|$ of the driving field on the qubit depends not only on the driving power $P_d$ but also on the frequency detuning $\omega_n-\omega_d$. In the dispersive region of the measured spectra where the sideband transitions appear, the driving frequency $\omega_d$ is largely detuned from the frequency $\omega_n$ of the resonator, i.e. $|\omega_n-\omega_d|\gg g_n$. The effective Rabi frequency $|\Omega_{q,n}|$ of the driving field on the qubit is much reduced compared to the Rabi frequency $\Omega_{r,n}$ of the driving field on the resonator, i.e., $|\Omega_{q,n}|\ll\Omega_{r,n}$.
In \fig{\ref{figs2}}, the effective Rabi frequencies $|\Omega_{q,n}|$ $(n=1,3)$ are plotted as a function of the corresponding frequency detunings $\omega_n-\omega_d$. Indeed, $|\Omega_{q,n}|$ is greatly reduced when $|\omega_n-\omega_d|\gg g_n$, indicating that the effective driving power on the qubit is extremely weak in this dispersive regime, as compared to the driving power originally applied to the resonator.

\subsection{Average photon number calibration}
\begin{figure}[!h]
\centering
\includegraphics[scale=1.1]{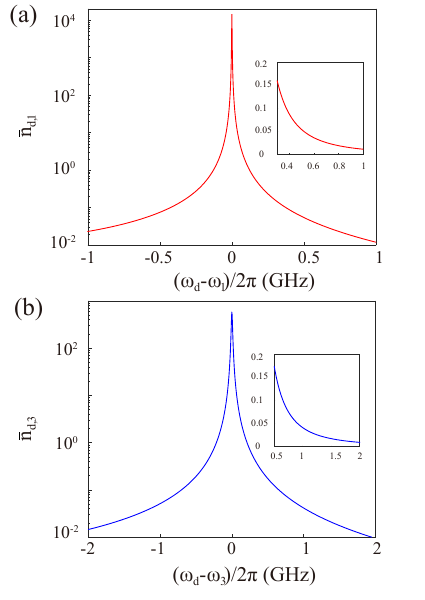}
\caption{
Average number of photons in the resonator that come from the driving field. (a) $\lambda/2$-mode and (b) $3\lambda/2$-mode. Each inset is a zoom of the off-resonance region between the driving field and the corresponding mode.}
\label{figs3}
\end{figure}

The average number of photons in the resonator that come from the driving field can be calculated from \cite{walls_quantum_2008,aspelmeyer_cavity_2014}
\bal
\bar{n}_{d,n} = \frac{\kappa_{n,{\rm ex}}/2}{{(\omega_n-\omega_d)}^2 + \kappa_n^2/4 }\frac{P_d}{\hbar \omega_d},
\label{eq_avgphd}
\eal
where we have used the same external loss rate for both sides of the resonator because they are nearly symmetric in our setup.
It is clear from \eq{\ref{eq_avgphd}} that the mean number $\bar{n}_{d,n}$ of the intracavity photons for the $n$th-mode, which are injected by the driving field, has a Lorentzian line shape centered around the frequency $\omega_n$ of the $n$th-mode with a width of $\kappa_{n,{\rm ex}}$. Therefore, although the driving field may provide thousands of photons into the resonator when it is on resonance with the resonator mode, the intracavity photon number $\bar{n}_{d,n}$ drops dramatically when the detuning $|\omega_n-\omega_d|$ is getting larger. We see from \fig{\ref{figs3}} that even for a small detuning (i.e., $|\omega_n-\omega_d|$ is on the order of $g_n$), the mean number of the intracavity photons injected by the driving field is already reduced to a single-photon level, and for a large detuning with $|\omega_n-\omega_d|\gg g_n$, the average photon number $\bar{n}_{d,n}$ is almost zero. It indicates that in our experiment the power of the driving filed is irrelevant to the high-order effects observed in the dispersive regime of the transmission spectra, where the negligible photons are injected by the driving field. This provides a clear and convincing evidence that the observed high-order processes are measured at the quantum limit of single or even fewer photon level.

\subsection{Schrieffer-Wolff transformation and higher-order couplings}
In the qubit eigenbasis $\{|g\rangle,\,|e\rangle\}$, the displaced Hamiltonian \eq{\ref{eq_hdfull}} can now be rewritten as
\bal
\label{eq_hdfull2}
\tilde{H}=&\frac{1}{2} \hbar\omega_q\sigma_z +\Omega_{d,q}\cos{\omega_d t}\, (\cos \theta \sigma_z -\sin \theta \sigma_x)\nonumber \\
&+\sum_{n=1,3} [\hbar\omega_n a_n^\dagger a_n +\hbar g_n (a_n^\dagger + a_n )(\cos \theta \sigma_z -\sin \theta \sigma_x)] \nonumber \\
\eal
which includes both a $\sigma_{z}$-component and a $\sigma_{x}$-component.
The $\sigma_{z}$-component in the pump Hamiltonian induces periodic oscillations of the qubit frequency, which produces effective qubit resonances at frequencies $\omega_{q}\mp s\omega_{d}$, with $s$ being an integer. When one of the effective resonances is near the frequency of a resonator mode, energy exchange between the qubit and the resonator will be enabled by the pump field. As a result, spectral lines in addition to the qubit frequency can be observed. The $\sigma_{x}$-component in the pump Hamiltonian flips the qubit state when the pump frequency is on resonance with the qubit frequency. Furthermore, when combined with (higher-order) qubit-resonator coupling, it can also induce high-order transitions at appropriate pump frequencies. Below we analyze the possible transitions in the qubit-resonator system in the dispersive limit with $\vert\Delta_{n}^{\pm}\vert\gg g_{n}$, where $\Delta_{n}^{\pm}=\omega_{q}\pm\omega_{n}$. In this limit, although the qubit and resonator modes cannot exchange energy directly due to frequency mismatch, higher-order processes can be induced by the Rabi coupling even for a very weak pumping field.

We apply a generalized Schrieffer-Wolff transformation to the Hamiltonian $ \tilde{H}$,
\bal
\tilde{H}_{\textrm{eff}} &= U^\dagger \tilde{H} U,
\label{eq_heff}
\eal
where the displacement operator $U$ is
\bal
U &=\exp{\Big\{{\sum_{n=1,3} \left[\lambda_{n,-} (\sigma_- a_n^\dagger - \sigma_+ a_n)  + \lambda_{n,+} (\sigma_- a_n - \sigma_+ a_n^\dagger)\right]}\Big \}},
\label{eq_u}
\eal
with $\lambda_{n,\pm} = -g_n \sin \theta/\Delta_{n}^{\pm}$ ~\cite{bravyi_schriefferwolff_2011,zueco_qubitoscillator_2009}. We then divide the Hamiltonian $\tilde{H}_{\textrm{eff}}$ into terms of different orders of the small parameter $\lambda_{n,\pm}$,
\bal
\tilde{H}_{\textrm{eff}}=H_{0} + \sum_{s}\tilde{H}^{(s)}.
\eal
In this expression,
\bal
H_{0}=\frac{1}{2}\hbar\omega_{q}\sigma_z + \sum_{n=1,3} (H_n + H_{\textrm{ac}}),
\eal
describes the uncoupled system Hamiltonian modified by the Stark and Bloch-Siegert shifts
\bal
H_{\textrm{ac}}=-\frac{\hbar}{2} \left[g_{n}\sin\theta (\lambda_{n,-}+\lambda_{n,+})(2a_{n}^{\dag}a_{n}+1)\sigma_{z}\right].
\eal
The term $\tilde{H}^{(s)}$ is to the $s$-th order of $\lambda_{n,\pm}$, which contains a time-dependent factor $\cos{\omega_d t}$ due to the pump field.

To gain more insight into the physics of the above higher-order terms, we now consider the Hamiltonian $\tilde{H}_{\textrm{eff}}$ in the interaction picture of $H_{0}$ with
\bal
\sigma_{\pm} \rightarrow \sigma_{\pm} e^{\pm i \omega_q t},~~a_n^\dagger \rightarrow a_n^\dagger e^{ i \omega_n t},~~a_n \rightarrow a_n e^{ -i \omega_n t}.
\eal
Below we study different order couplings under the rotating-wave approximation by neglecting all fast oscillating terms.

{\bf Zeroth-order term $\tilde{H}^{(0)}$.} With a pump frequency $\omega_{d}=\omega_{q}$ and under the rotating-wave approximation, the zeroth-order term becomes
\bal
\tilde{H}^{(0)}= X_{n}^{(0)} \sigma_{x},
\eal
with $X_{n}^{(0)} = -\Omega_{d,q}\sin \theta/2$, which generates a transition between the two qubit states. This transition gives the measured qubit spectrum in Fig. 2.

{\bf First-order term $\tilde{H}^{(1)}$.} At the pump frequency $\omega_{d}=\omega_{q}\mp\omega_{n}$, the driving on the $\sigma_z$-component of the qubit yields an effective Hamiltonian describing the first-order red (``$-$'') or blue (``$+$'') sideband transitions.

Another contribution to the first-order couplings occurs at the pump frequency $\omega_{d}=\omega_{n}$. Here the driving on the $\sigma_{x}$-component of the qubit, together with the qubit-resonator coupling, produces an indirect driving on the resonator modes. The effective driving on the resonator then reads
\bal
H_{\textrm{eff}} = Z_n^{(1)} (a_{n}+a_{n}^{\dag})\sigma_{z},
\eal
with
\bal
Z_n^{(1)}= \frac{\Omega_{d,q}}{2}\sin\theta (\lambda_{n,-} +\lambda_{n,+}).
\eal

{\bf Second-order term $\tilde{H}^{(2)}$.} The $\sigma_{z}$-component driving generates second-order terms in the forms of
\bal
H_{\textrm{eff}} = Z_n^{(2)}(a_{n}^{2} + a_{n}^{\dag 2})\sigma_{z},
\eal
with
\bal
Z_n^{(2)} = - \Omega_{d,q}\cos\theta\lambda_{n,-}\lambda_{n,+}
\eal
at the pump frequency $\omega_{d}=2\omega_{n}$, which are two-photon processes. The $\sigma_{z}$-component also generates cross-mode coupling terms between different resonator modes in the forms of
\bal
H_{\textrm{eff}} = Z_{\bar{1}3}^{(2)} (a_{3}^{\dag}a_{1} + a_{1}^{\dag}a_{3})\sigma_{z},
\eal
with
\bal
Z_{\bar{1}3}^{(2)} = -\Omega_{d,q} \cos\theta(\lambda_{1,-}\lambda_{3,-}+\lambda_{1,+}\lambda_{3,+}),
\eal
at $\omega_{d}=\omega_{3}-\omega_{1}$, and
\bal
H_{\textrm{eff}} =Z_{13}^{(2)} (a_{3}^{\dag}a_{1}^{\dag} + a_{1}a_{3})\sigma_{z},
\eal
with
\bal
Z_{13}^{(2)}=- \cos\theta(\lambda_{1,-}\lambda_{3,+}+\lambda_{1,+}\lambda_{3,-})
\eal
at $\omega_{d}=\omega_{3}+\omega_{1}$, respectively. These terms cause effective couplings between different resonator modes with the frequency difference compensated by the pump frequency.

The $\sigma_{x}$-component driving generates second-order coupling terms between the qubit and the resonator modes, which are responsible for the second-order single-photon-driven sideband transitions observed in the measurement.
Similarly, higher-order terms can be analyzed. In particular, at the frequency $\omega_{d}=\omega_{q}-3\omega_{1}$, a third-order single-photon-driven red-sideband transition can be generated.  Discussions on these sideband terms will be straightforward using our approach.

\subsection{Sideband transitions}
In the above, we have analyzed possible higher-order terms induced by the pump field and the qubit-resonator coupling. Here we discuss dominant contributions among all terms that are directly connected to the measured red and blue-sideband transitions in the experiment.

First, at pump frequencies $\omega_d= \omega_q -s \omega_n$ $(s=1,2,3)$, the interaction terms that are not rapidly oscillating are described by the effective Hamiltonian
\bal
H_{n,{\textrm{red}}}^{(s)} &= R_{n}^{(s)} \left( {a_n^\dagger}^s \sigma_- + a^s_n \sigma_+ \right),
\label{eq_red}
\eal
with the coefficients
\bal
R_{n}^{(1)} &= -\Omega_{d,q} \cos \theta \lambda_{n,-}, \\
R_{n}^{(2)} &= -\frac{\Omega_{d,q}}{2} \sin \theta \,\lambda_{n,-}(\lambda_{n,-}+\lambda_{n,+}), \\
R_{n}^{(3)} &= \frac{2\Omega_{d,q}}{3} \cos \theta \,\lambda_{n,-}^2 \lambda_{n,+}. \label{R3}
\eal
We denote these terms as the $s$th-order single-photon-driven red-sideband transition for the $n\lambda/2$-mode. These terms generate exchanges between the states $\ketc{g,N_n}$ and $\ketc{e,N_n-s}$ with amplitudes $\propto \lambda_{n,\pm}^{s}$, which can effectively convert a qubit excitation into $s$ photon excitations of frequency $\omega_n$. Note that if there are no counter-rotating terms in the Hamiltonian, $\lambda_{n,+}=0$ in the displacement operator $U$ given in Eq.~(\ref{eq_u}). Thus, the third-order process related to the nonzero coefficient in Eq.~(\ref{R3}) is due to the existence of the counter-rotating terms.

Similarly, when driving at $\omega_d= \omega_q +s\omega_n$ $(s=1,2,3)$, we obtain single-photon-driven blue-sideband transition for the $n\lambda/2$-mode with
\bal
H_{n,{\textrm{blue}}} &= B_n^{(s)} \left( {a_n^\dagger}^s \sigma_+ + a^s_n \sigma_- \right),
\label{eq_blue}
\eal
where the coefficients are
\bal
B_{n}^{(1)} &= -\Omega_{d,q} \cos \theta \lambda_{n,+}, \label{B1} \\
B_{n}^{(2)}  &= -\frac{\Omega_{d,q}}{2} \sin \theta \lambda_{n,+}(\lambda_{n,-}+\lambda_{n,+}), \label{B2} \\
B_{n}^{(3)}  &= \frac{2\Omega_{d,q}}{3} \cos \theta \, \lambda_{n,-}\lambda_{n,+}^2. \label{B3}
\eal
It couples the states such as $\ketc{g,N_n}$ and $\ketc{e,N_n+s}$ and produces a qubit excitation and $s$ photon excitations in the $n\lambda/2$-mode simultaneously. Obviously, the blue-sideband transitions related to the nonzero coefficients in Eqs.~(\ref{B1})-(\ref{B3}) are also due to the existence of the counter-rotating terms in the Hamiltonian, because these coefficients are proportional to either $\lambda_{n,+}$ or $\lambda_{n,+}^2$.

Interestingly, within our measured spectral range, the single-photon-driven second-order terms also include a cross-mode red-sideband transition when driving at $\omega_d= \omega_q \pm \omega_1 - \omega_3$ with the coupling Hamiltonian
\bal
H_{c,{\textrm{red}}}^{(2)} &= \sigma_- {a_3^\dag}\left( R^{(2)}_{\bar{1}3} \,a_1 + R^{(2)}_{13} \,{a_1^\dag} \right)+ { \rm h.c.} ,
\label{eq_cred}
\eal
and a single-photon-driven cross-mode blue-sideband transition when driving at $\omega_d= \omega_q \pm \omega_1 + \omega_3$
\bal
H_{c,{\textrm{blue}}}^{(2)} &= \sigma_- {a_3} \left( B^{(2)}_{1\bar{3}} \, a_1^\dag  +  B^{(2)}_{\bar{1}\bar{3}}{a_1}\right) + { \rm h.c.} .
\label{eq_cblue}
\eal
The coupling coefficients for these cross-mode sideband transitions are
\bal
R^{(2)}_{\bar{1}3} &= - \frac{\Omega_{d,q}}{2} \sin \theta \,[\lambda_{1,+}(\lambda_{3,-}+\lambda_{3,+}) + \lambda_{3,-}(\lambda_{1,-}+\lambda_{1,+})], \nonumber \\
R^{(2)}_{13} &= - \frac{\Omega_{d,q}}{2} \sin \theta \,[\lambda_{1,-}(\lambda_{3,-}+\lambda_{3,+}) + \lambda_{3,-}(\lambda_{1,-}+\lambda_{1,+})],\nonumber \\
B^{(2)}_{1\bar{3}} &= - \frac{\Omega_{d,q}}{2} \sin \theta \,[\lambda_{1,-}(\lambda_{3,-}+\lambda_{3,+}) + \lambda_{3,+}(\lambda_{1,-}+\lambda_{1,+})], \nonumber \\
B^{(2)}_{\bar{1}\bar{3}} &= - \frac{\Omega_{d,q}}{2} \sin \theta \, [\lambda_{1,+}(\lambda_{3,-}+\lambda_{3,+}) + \lambda_{3,+}(\lambda_{1,-}+\lambda_{1,+})].
\eal
The single-photon-driven cross-mode red-sideband transitions are $\ketc{e10} \leftrightarrow \ketc{g01}$ and $\ketc{e00} \leftrightarrow \ketc{g11}$. The single-photon-driven cross-mode blue-sideband transitions are $\ketc{e11} \leftrightarrow \ketc{g00}$ and $\ketc{e01} \leftrightarrow \ketc{g10}$. Note that other cross-mode sideband transitions could appear if we include higher-order terms in the effective Hamiltonian.

The amplitude of the $s$th-order sideband transition depends on $\lambda_{n,\pm}^{s}$ with $\lambda_{n,\pm} \ll 1$, and thus it decreases very quickly as the order $s$ increases. For illustration, we choose $\omega_{q}/2\pi=6.85$ GHz and $\omega_{q}/2\pi=14.53$ GHz to calculate the coupling coefficients for the first mode and the third mode, respectively, as shown in Table I. Here the first-order transitions have normalized amplitudes $\sim 1\times 10^{-2}$ and the amplitudes of the third-order transitions decrease to $\sim 1\times 10^{-5}$.
\begin{widetext}

\begin{table}[h!]
\centering
\caption{Sideband transition coefficients normalized to the driving strength $\Omega_{d,q}$.}
\label{tab1}
\begin{tabular}{|c||c|c|c|c|c|c|c|c|c|}
 \hline\hline
$\omega_{q}/2\pi \textrm{(GHz)}$ &$X_{1}^{(0)}$ &$R_{1}^{(1)}$ &$R_{1}^{(2)}$ &$R_{1}^{(3)}$ & $B_{1}^{(1)}$&  $B_{1}^{(2)}$&  $B_{1}^{(3)}$ &$ R^{(2)}_{\bar{1}3}$&  $B^{(2)}_{1\bar{3}}$\\
\hline
 6.85 &$-4.4\times10^{-1}$   & $3.5\times10^{-2}$   & $-3.1\times10^{-3}$  &$-4.5\times10^{-5}$ & $1.3\times10^{-2}$  & $-1.2\times10^{-3}$& $-1.7\times10^{-5}$ & $9.5\times10^{-3}$& $3.5\times10^{-3}$\\
\hline\hline
$\omega_{q}/2\pi \textrm{(GHz)}$ &$X_{3}^{(0)}$ &$R_{3}^{(1)}$ &$R_{3}^{(2)}$ &$R_{3}^{(3)}$ & $B_{3}^{(1)}$&  $B_{3}^{(2)}$&  $ B_{3}^{(3)}$ &$R^{(2)}_{13}$&  $B^{(2)}_{\bar{1}\bar{3}}$\\
\hline
14.53 &$-2.1\times10^{-1}$ &$3.8\times10^{-2}$  & $-4.4\times10^{-4}$ &$-9.7\times10^{-6}$ & $8.2\times10^{-3}$ &$-9.5\times10^{-5}$ & $-1.6\times10^{-6}$ &$-2.8\times10^{-4}$ & $-1.1\times10^{-4}$\\
 \hline\hline
\end{tabular}
\end{table}
\end{widetext}

%\end{widetext}

\vspace{8pt}
\end{document}